# Cluster-based Random Radial Basis Kernel Function for Hyperspectral Data Classification

**Niazmardi, Saeid**

**Department of Surveying Engineering, Faculty of Civil and Surveying Engineering, Graduate University of Advanced Technology**

**s.niazmardi@kgut.ac.ir**

**Abstract**

Kernel-based classification methods, particularly the support vector machine (SVM), are among the most common algorithms for hyperspectral data classification. The Radial Basis function (RBF) kernel has earned great popularity in hyperspectral data classification due to its superior performance among other available kernel functions. Nonetheless, the cross-validation technique usually used for tunning the RBF parameter can be time-consuming and may result in sub-optimal values for the parameter. This paper proposed the cluster-based random radial basis function (CRRBF) kernel function as an alternative to the RBF kernel to achieve similar performance with a more manageable parameter, which is the number of clusters. The CRRBF kernel initially clusters the hyperspectral bands and then constructs an RBF kernel with a randomly assigned value as the kernel parameter from each cluster of bands. The final CRRBF kernel is constructed by adding up these basis RBF kernels. We have designed several experiments to evaluate the SVM performance trained with the CRRBF kernel considering a different number of clusters and training samples, using three hyperspectral data sets. The obtained results showed that the CRRBF kernel could provide comparable or better results than the RBF. The results also showed that the classification performance is pretty robust to the number of clusters, as the only open parameter of the CRRBF kernel.

# 1. Introduction

Thanks to its high spectral resolution, hyperspectral data can provide more spectral information about the scene than other remote sensing data modalities [1]. Along with this potential, hyperspectral data comes with specific challenges intrinsic to its nature. Some of them, such as high dimensionality and limited labeled samples, pose serious challenges to hyperspectral data classification. Accordingly, several machine learning algorithms have been proposed and used for hyperspectral data classification [1-3].

Deep learning and kernel-based methods are two common categories of algorithms for hyperspectral data classification. Compared with deep learning, kernel methods are more effective with small samples and medium or small-sized datasets [4, 5]. Thus, over the last two decades, kernel-based classification methods, exemplified by the support vector machine (SVM), have been turned into benchmark algorithms for hyperspectral data classification [6, 7]. The main idea of kernel-based methods is finding a simpler (more linear) data representation using kernel functions [8]. Kernel functions implicitly specify an inner product in a high-dimensional Hilbert space, called feature or kernel space, in which the SVM algorithm seeks a separating hyper-plane with the maximum margin [9-11].

The kernel methods are very efficient regarding training time and performance [12]. These algorithms also have elegant compatibility with data dimensionality and limited training samples [13]. Despite having a solid theoretical background, the performance of kernel-based algorithms is tied to properly choosing a kernel function and tuning its parameters.

To use a kernel-based algorithm, the user should choose the kernel function from a multitude of kernel functions that satisfy the mercer's conditions [14]. The most common choices are linear, polynomial, and Radial Basis Functions (RBF) [15]. Among all these kernel functions, the RBF kernel is the most popular one, due to its outstanding performance.

However, several other alternative kernels have been proposed for classification, such as the Chebyshev kernel [16], Legendre kernel [17], Hermite kernel [18], Lie group kernel [19], and Random forest kernel [20]. Despite yielding acceptable classification performances, these alternative kernels have two problems. First, they also have some free parameters which affect the classification performance. Second, their performances are comparable with the performance of the RBF kernel. Thus, their applications are only limited to particular cases.

The problem of tuning the kernel parameter is usually addressed using the K-fold cross-validation (CV) technique. Using an exhaustive search strategy; the CV technique searches a set of predefined candidate values for the kernel parameter to find the best-performing one. Despite having good performance for parameter selection [21-23], adapting the exhaustive search strategy increases the computational time of the CV technique. Besides, its performance is affected by the size and representativeness of the training set [24].

Different search strategies have been proposed to substitute the exhaustive search of the CV technique and increase its performance in terms of computational time or classification accuracy. Among these strategies, meta-heuristic search methods such as the genetic algorithm [25], particle swarm optimization [26, 27], and immune system [28] are the most used ones.

The other approach to kernel parameter selection is adapting parameter-free kernels. Ding et.al proposed a Random RBF (RRBF) kernel in [29, 30]. The RRBF kernel for $d$-dimensional data is calculated as the sum of $d$-RBF kernels, each constructed using one dimension of the data, considering a random value as its parameter [30]. Despite its promising classification performance, The RRBF kernel cannot properly handle the hyperspectral data; since the similar information content of the adjacent bands will negatively affect its performance.

Inspired by the superior performance of the RBF kernel and the parameter-free RRBF kernel, we proposed a novel kernel function called the cluster-based random RBF (CRRBF) kernel to address the problem of kernel parameter selection for hyperspectral data classification. The CRRBF kernel initially clusters the hyperspectral bands into a predefined number of clusters and then uses a randomly assigned parameter to construct an RBF kernel from each cluster of bands. The final CRRBF kernel is constructed by summing these RBF kernels. Although the CRRBF kernel is not parameter-free, its only parameter is the number of clusters, which is much more manageable than the original RBF kernel parameter.

In the rest of this article, the RRBF kernel, and the proposed CRRBF kernels are presented in the second section. The third section introduces the used data sets and the experimental setups for evaluating the performances of the proposed kernel. The obtained results are presented and discussed in the fourth section, and we draw our conclusion in the fifth section.

2. **Methodology**

This section introduces the RRBF kernel and the proposed CRRBF kernel.

### 2.1. Random Radial Basis Kernel

The RBF kernel is defined as:

$$\mathbf{K}(\mathbf{x},\mathbf{y}) = \exp\left(-\gamma \|\mathbf{x}-\mathbf{y}\|^2\right) \quad (1)$$

Where $\gamma$, as the only open parameter in this kernel, controls its performance.

The RBF kernel is quite popular in remote sensing data analyses due to having low computational cost, flexibility, good learning capability, and supporting complex models [31, 32]. Nevertheless, the selected value for $\gamma$ highly affects the performance of the kernel-based learning algorithms that adapt this kernel.

Ding et.al. addressed the issue of RBF parameter selection by proposing a modified version of the RBF kernel called the RRBF [29, 30]. For two *d*-dimensional vectors $\mathbf{x}=(x_1, x_2, ..., x_d)^T$ and $\mathbf{y}=(y_1, y_2, ..., y_d)^T$, The RRBF kernel is estimated as:

$$K(\mathbf{x},\mathbf{y}) = \exp\left(-\sum_{i=1}^{d}\gamma_i (x_i - y_i)^2\right) \quad (2)$$

Unlike the RBF kernel that uses a single parameter, the RRBF kernel considers separate parameters, shown by $\gamma_i, i=1,2,...,d.$ for each dimension of the data. The RRBF kernel assigns *d* random values to these parameters without considering the training data. Doing so, the RRBF kernel tries to be a parameter-free version of the RBF kernel.

### 2.2. Cluster-based Random Radial Basis Kernel

Despite having acceptable performance, the RRBF kernel cannot yield acceptable results in the case of using high-dimensional data such as hyperspectral data sets. Because the information content of some bands of hyperspectral data is irrelevant to the learning task. Due to random assignment of the weights, these bands may be assigned to large weights which have an adverse effect on the representation capability of the kernel and the learning task. To address this issue, we proposed the CRRBF. This kernel initially clusters the available spectral bands of the data into *k* clusters $C_1, C_2, ..., C_k$, using a partitioning algorithm such as K-Means. Then, an RBF kernel with a random value is calculated for each cluster of bands. Finally, these RBF kernels with random parameters are added up to construct the final kernel. The formula for calculating the CRRBF kernel is as follows:

$$K(\mathbf{x},\mathbf{y}) = \exp\left(-\sum_{C_k=1}^{k}\sum_{i \in C_k}\gamma_k (x_i - y_i)^2\right) \quad (3)$$

This kernel only uses *k* random number as its parameter $\gamma_k$. Thus, the CRRBF behaves less randomly than the RRF kernel. Besides, due to constructing the RBF kernels from clusters of bands, the CRRBF kernel can handle the data dimensionality.

## 3. Data Sets and Experimental Setup

### 3.1. Data sets

The proposed CRRBF has been evaluated for classifying three hyperspectral datasets. The first data set is the Pavia University data set, acquired by reflective optics system imaging spectrometer (ROSIS-3) over the Pavia University, Northern Italy, in 2001, with a spatial resolution of 1.3 m. This data set has 103 spectral bands and contains nine classes whose labeled data has been divided into 3921 training samples and 42776 test samples.

The second data set was acquired by the Airborne Visible/Infrared Imaging Spectrometer (AVIRIS) sensor over the Kennedy Space Center (KSC), Florida, on March 23, 1996. This dataset has 18m spatial resolution and 176 spectral bands. From 13 classes of this data set, 5211 labeled samples were extracted. For the experiments of this paper, 40 percent of the available labeled samples were randomly selected as training samples, and the remaining 60 percent were considered test samples.

The third data set was gathered over the University of Houston campus and the neighboring urban area using the CASI sensor. This hyperspectral data set has 144 spectral bands and a spatial resolution of 2.5 m. The labeled samples for 15 classes of this data set have been divided into 2382 training samples and 12197 test samples.

Table 1 lists the number of classes and their respective number of labeled samples. Figure 1 shows the true color composite of the used data sets.

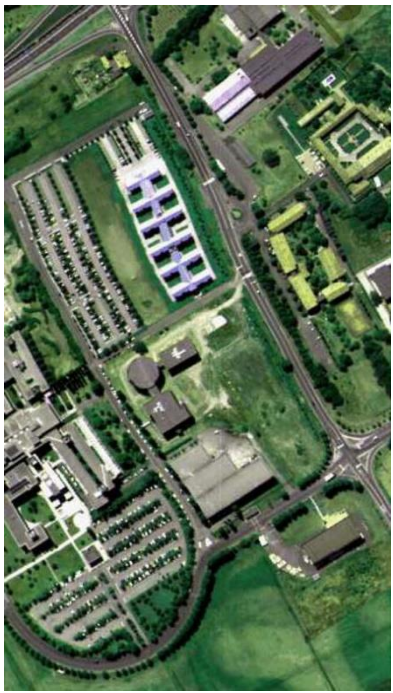
(a)

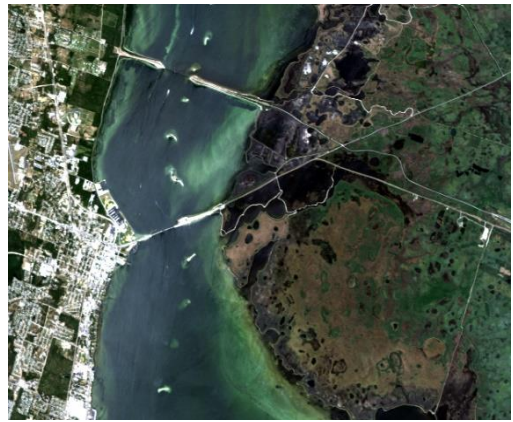
(b)

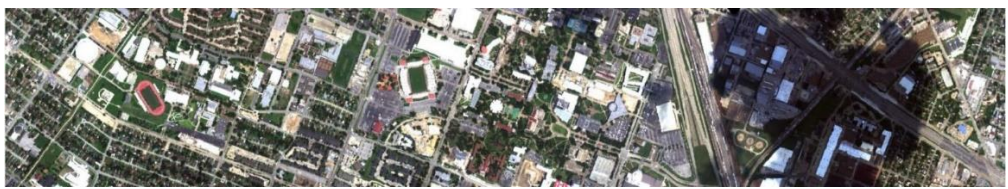
(c)

**Figure 1**. The true color composite image of the used hyperspectral data sets, (a) University of Pavia data set, (b) Kennedy space center data set, and (c) University of Huston data set

Table 1. Classes and number of samples labeled for training and test in the used hyperspectral data sets

| University of Pavia | | | Kennedy Space center | | | University of Huston | | |
|---|---|---|---|---|---|---|---|---|
| Class name | Number of samples | | Class name | Number of samples | | Class name | Number of samples | |
| | Train | Test | | train | Test | | Train | Test |
| Asphalt | 548 | 6641 | Scrub | 304 | 457 | Healthy grass | 198 | 1053 |
| Meadows | 540 | 18649 | Willow swamp | 97 | 146 | Stressed grass | 190 | 1064 |
| Gravel | 392 | 2099 | Cabbage palm hammock | 102 | 154 | Synthetic grass | 192 | 505 |
| Trees | 524 | 3064 | Cabbage palm/Oak hammock | 100 | 152 | Tree | 188 | 1056 |
| Metal Sheets | 265 | 1345 | Slash pine | 64 | 97 | Soil | 186 | 1056 |
| Soil | 532 | 5029 | Oak/broadleaf hammock | 91 | 138 | Water | 182 | 143 |
| Bitumen | 375 | 1330 | Hardwood swamp | 42 | 63 | Residential | 196 | 1072 |
| Bricks | 514 | 3680 | Graminoid marsh | 172 | 259 | Commercial | 191 | 1053 |
| Shadows | 231 | 947 | Spartina marsh | 208 | 312 | Road | 193 | 1059 |
| Total | 3921 | 42776 | Cattail marsh | 161 | 243 | Highway | 191 | 1036 |
| | | | Salt marsh | 167 | 252 | Railway | 181 | 1054 |
| | | | Mud flats | 201 | 302 | Parking lot1 | 192 | 1041 |
| | | | water | 307 | 557 | Parking lot2 | 184 | 285 |
| | | | | | | Tennis court | 181 | 247 |
| | | | | | | Running track | 187 | 473 |

### 3.2. Experimental setup

We designed several experimental scenarios to evaluate the classification performance of the SVM algorithm trained with the proposed kernel function. To analyze the role of the number of clusters, as the only free parameter of the CRRBF, we estimate the classification accuracy of an SVM algorithm trained with the proposed kernel considering 3, 4, 5, 6, 7, 8, 9, and 10 clusters. In this scenario, the trade-off parameter of the SVM algorithm was selected using the exhaustive search strategy from the set {1, 2, 4, 8, 16, 32, 64, 128, 256, 512, 1024}. we also trained an SVM algorithm using the CRRBF considering 10, 20, 30, 40, 50, and 75 percent of the available training samples for each class of considered datasets, to study the performance of the proposed kernel using a limited number of training samples. In this scenario, the number of clusters and the SVM trade-off value were set to the values that yielded the best performances in the first classification scenario. To avoid biased conclusions as a result of the random nature of the CRRBF, each classification scenario that used this kernel function was repeated ten times, and the average value of classification accuracies was reported. In all the implementations of CRRBF kernel, the K-Means algorithm is adapted as the clustering algorithm.

For comparison, we trained the SVM algorithms by adapting RRBF, RBF, and polynomial kernel functions. A 5-fold cross-validation technique was used to tune the RBF parameter, polynomial degree, and the SVM trade-off parameter. The RBF parameter was selected from the values between 0.01 and 20 with a step size of 1, the Polynomial degree was selected from the set of integers less than 11, and the SVM trade-off was selected from {1, 2, 4, 8, 16, 32, 64, 128, 256, 512, 1024}.

## 4. Results and Discussion

### 4.1. Analyzing the sensitivity of CRRBF kernel to the parameters

Figure 2 shows the average overall accuracy of an SVM algorithm trained with ten CRRBF kernels, using different values as the SVM trade-off parameter.

As observable in this Figure, the SVM algorithm yielded acceptable classification performances in all the data sets. It is also observable that the classification results are more affected by the value used as the SVM trade-off than they are affected by the number of clusters.

To better demonstrate the role of the number of clusters on the performance of the proposed kernel function, Table 2 lists the highest obtained accuracies considering different values for trade-off parameters for each number of clusters.

**Table 2**. The maximum obtained accuracy of SVM algorithm training with ten CRRBF considering different number of clusters

| Data set | Number of clusters | | | | | | | |
|---|---|---|---|---|---|---|---|---|
| | 3 | 4 | 5 | 6 | 7 | 8 | 9 | 10 |
| **University of Pavia** | 81.44 | 81.94 | 82.44 | 81.03 | 81.69 | 81.41 | 81.73 | **82.46** |
| **University of Huston** | 78.63 | 78.64 | **79.39** | 78.87 | 78.51 | 78.64 | 77.96 | 78.45 |
| **Kennedy space center** | **95.20** | 95.14 | 95.05 | 95.07 | 95.06 | 94.72 | 95.09 | 95.09 |

The results show that the proposed kernel is very robust to the selected value as the number of clusters. The standard deviation of the overall accuracies of the classification using ten CRRBF kernels were 0.47, 0.38, and 0.13 for University of Pavia, University of Huston, and Kennedy space center datasets, respectively. Table 3 tabulates the SVM trade-off value and the cluster numbers using which the SVM algorithm yielded the highest classification accuracy.

**Table 3**. Parameters that yielded the highest classification accuracy

| Data set | Classification accuracy | | Parameter values | |
|---|---|---|---|---|
| | Overall accuracy | Kappa | Number of clusters | Trade-off parameter |
| **University of Pavia** | 82.46 | 0.78 | 10 | 512 |
| **University of Huston** | 79.39 | 0.78 | 5 | 1024 |
| **Kennedy space center** | 95.20 | 0.95 | 3 | 512 |

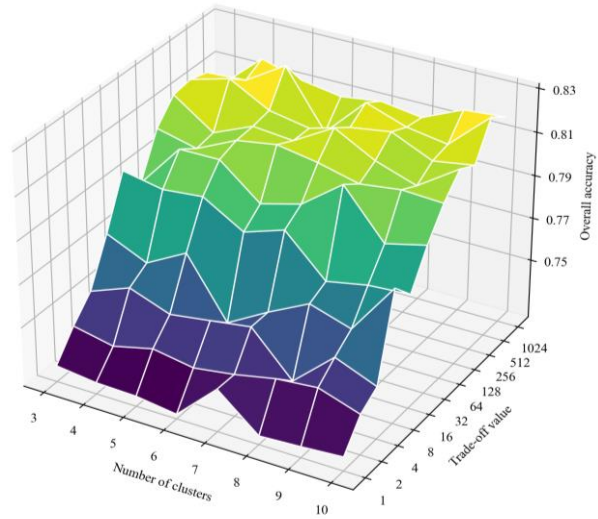

(a)

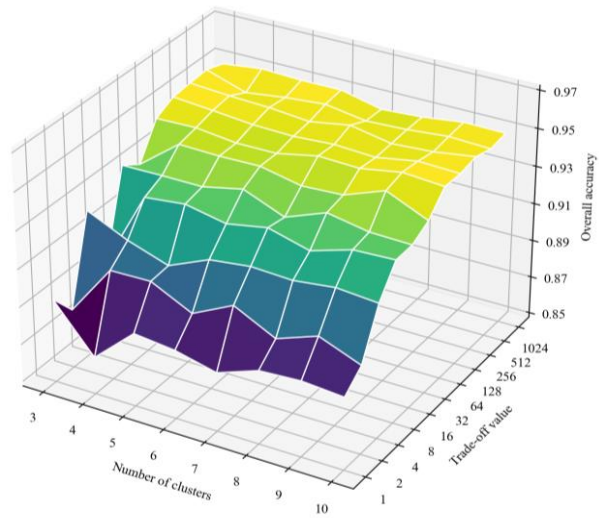

(b)

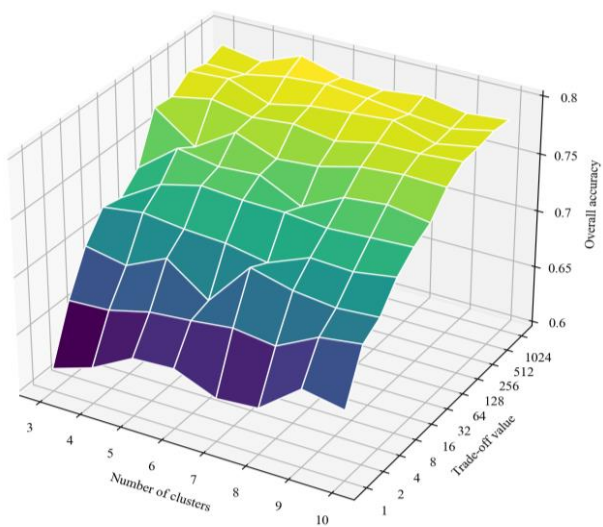

(c)

**Figure 2**. The Average overall accuracy of classification considering different values for cluster numbers and trade-off parameters for a) University of Pavia, b) Kennedy space center data sets, and C) University of Huston.

## 4.2. Analyzing the sensitivity of CRRBF kernel to the number of training samples

Figure 3 shows the performance of an SVM algorithm trained using the proposed CRRBF kernel considering sets with different numbers of training samples. Based on the results, the proposed CRRBF kernel was able to provide acceptable performances, even in the case of using 10 percent of training samples to train the SVM algorithm. For example, the average overall accuracy of the classification was 75.34, 86.90, and 71.17 respectively for University of Pavia, Kennedy space center, and university of Huston datasets, in the case of using 10 percent of training samples.

The large variation of classification accuracies in the case of using a few numbers of samples for training can be attributed to the limited ability of a small number of samples to represent classes in the data sets.

## 4.3 Comparison with other kernel functions

The obtained results from an SVM algorithm trained using different kernels are presented in Figure 4. Apart from RBF and polynomial kernels, whose parameters are tunned using 5-fold cross-validation, this Figure also represents the average accuracy of the SVM trained using ten CRRBF and RRBF kernels.

As the results show, the CRRBF kernel was able to provide comparative and, in some cases, such as University of Huston data set, even better classification accuracies than the RBF kernel. The accuracy obtained from adapting the polynomial kernel was marginally less than the accuracies obtained from the RBF kernel. And the RRBF kernel yielded the lowest accuracies for all the data sets.

Since the adjacent spectral bands of hyperspectral data have similar information contents, using different random values for each spectral band will result in different kernels, some of which cannot properly represent the information class of data. Accordingly, adding up several kernels with low representation capabilities will result in low classification accuracies. The CRRBF kernels can overcome this issue by constructing the kernels from a few clusters of bands. These clusters have different information content and assigning random values as the kernel parameter will have limited effects on the representing capability of the basis RBF kernels.

Although the polynomial and RBF kernels were able to yield acceptable results, tunning their parameters using the CV technique is very time-consuming. The processing times for a 5-fold CV technique using different numbers of training samples are listed in Table 4. These processing times were obtained using MATLAB on a standard laptop PC ruining an Intel Core i5, 2.70 GHZ CPU, and 16 GB of RAM.

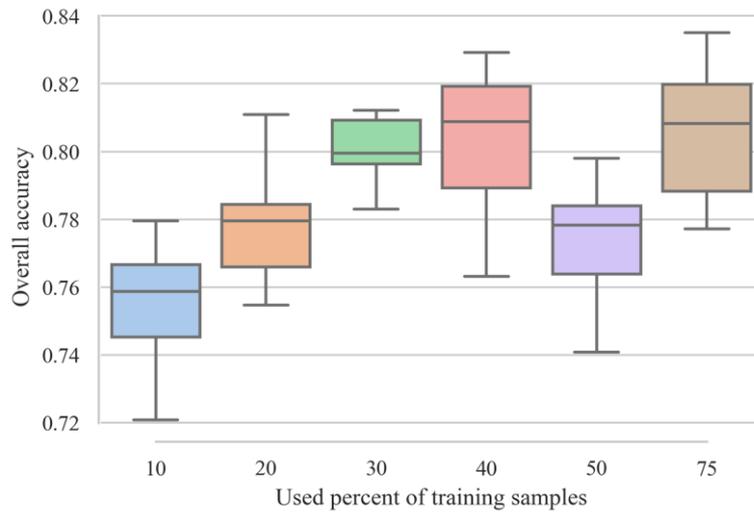

(a)

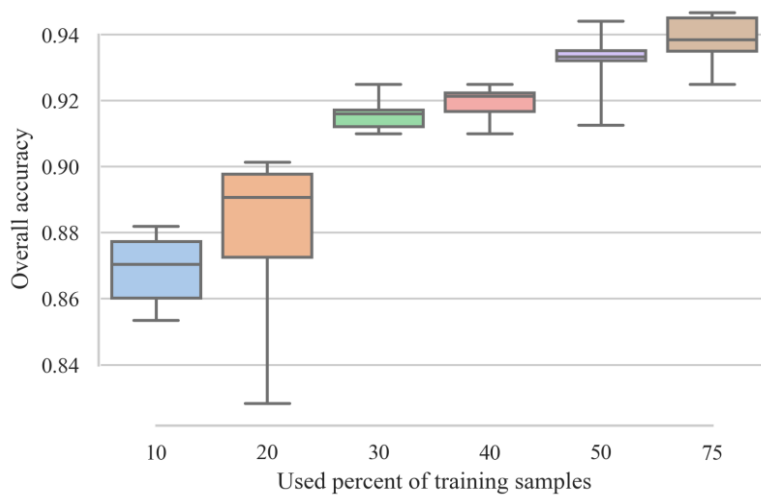

(b)

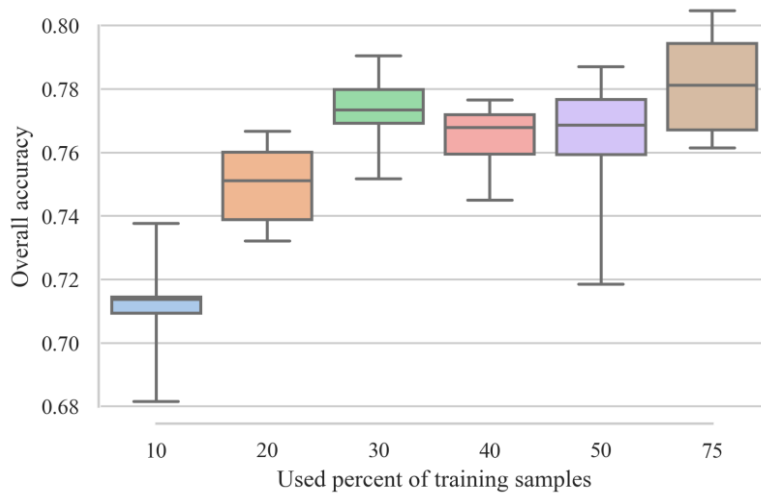

(c)

**Figure 3**. The overall accuracy of the SVM trained with a limited number of training samples while adapting ten CRRBF kernels, a) University of Pavia, b) Kennedy space center data sets, and C) University of Huston.

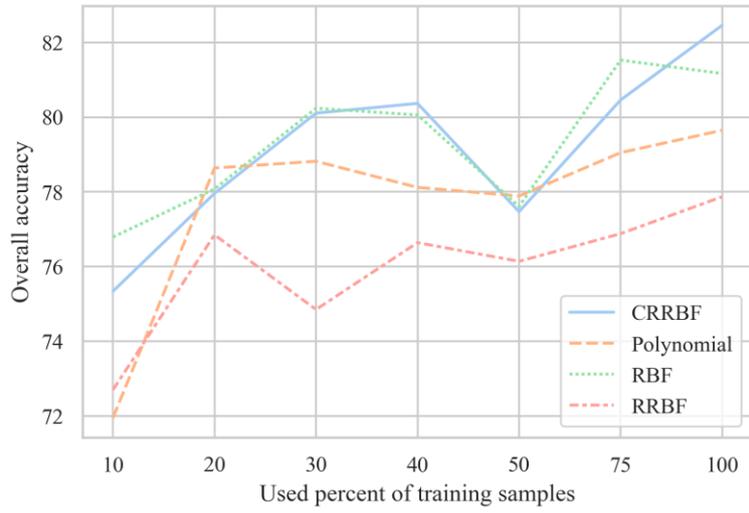

(a)

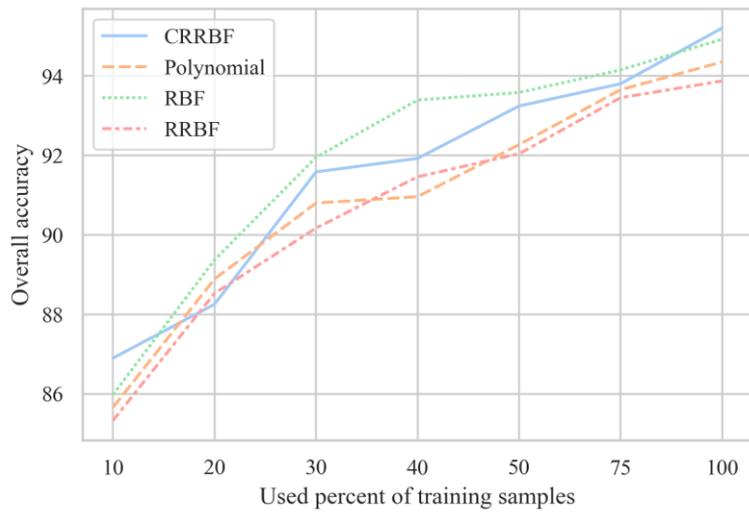

(b)

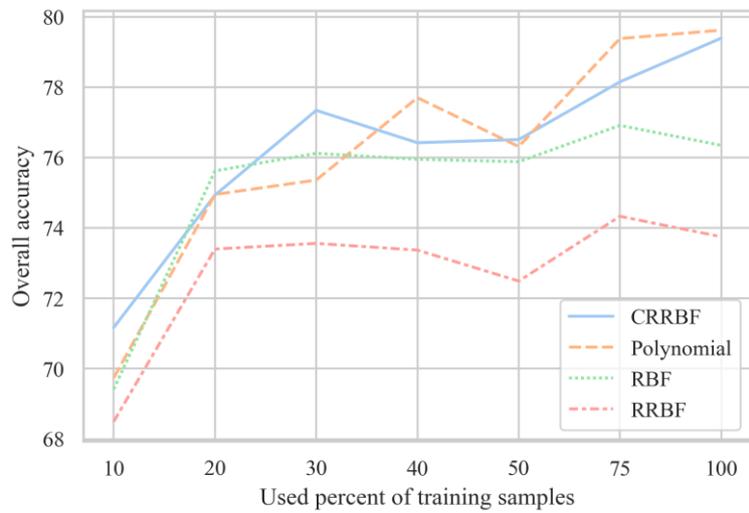

(c)

**Figure 4.** The overall accuracy of the SVM adapting different kernel functions and trained with different numbers of training samples, a) University of Pavia, b) Kennedy space center data sets, and C) University of Huston.

| Data set | Kernel functions | 10 | 20 | 30 | 40 | 50 | 75 | 100 |
|---|---|---|---|---|---|---|---|---|
| **Pavia university** | **RBF** | 25.5 | 77.0 | 116.2 | 222.2 | 296.0 | 564.9 | >1000 |
| | **Polynomial** | 48.0 | 204.6 | 586.7 | >1000 | >1000 | >1000 | >1000 |
| **KSC** | **RBF** | 15.5 | 54.5 | 115.7 | 184.4 | 271.0 | 547.20 | 764.8 |
| | **Polynomial** | 2.2 | 5.3 | 10.7 | 20.1 | 24.6 | 52.3 | 87.5 |
| **Huston university** | **RBF** | 20.6 | 55.4 | 101.5 | 140.1 | 196.1 | 367.8 | 523.2 |
| | **Polynomial** | 11.0 | 43.5 | 82.1 | 135.7 | 190.5 | 359.6 | 487.8 |

**Table 4.** The processing time (in seconds) using 5-fold cross-validation

As observable in this Table, the processing time depends on the number of used training samples. For example, the 5-fold CV techniques took more than 1000s to tune the RBF kernel for University of Pavia data set with 40 percent of training samples. The fact that the CRRBF kernel does not need CV-technique for tuning its parameter, justifies its use for classification of hyperspectral data.

## 5. Conclusions

Kenel-based classification algorithms are among the most successful algorithms for the classification of hyperspectral data. However, it is well-established that the performance of kernel-based learning algorithms is highly affected by the used kernel functions and the values used as the kernel parameters. The RBF kernel is one of the best-performing kernels for the classification of hyperspectral data, whose parameter is usually tunned using the n-fold CV technique. Nevertheless, the CV technique is usually a time-consuming technique that may lead to sub-optimal values for the kernel parameter. To have a kernel function with similar performance to the RBF kernel and avoid spending extra time on the tuning RBF parameter, we proposed a cluster-based random radial basis function kernel (CRRBF). This kernel function clusters the hyperspectral bands into a pre-defined number of clusters and then constructs an RBF kernel from each cluster of bands, considering a random value as the kernel parameter. Then these kernels are summed-up to construct the final kernel. We can draw the following conclusions based on the obtained results from our experiments:

- The CRRBF kernel can provide comparative and even better results than the RBF kernel. Thus, it can be used as a substitute for the RBF kernel for hyperspectral data classification.
- The number of clusters is the only open parameter of the CRRBF kernel. However, the classification results were quite robust to this parameter value. Besides, this parameter, as a result of being an integer, has a much-limited range than the RBF parameter.
- The CV technique that is usually used for parameter tuning of RBF and polynomial kernel can be very time-consuming. Our results showed that this step can be omitted in the case of adapting the CRRBF kernel.

For future studies, we plan to investigate how using the weighted summation of the RBF kernels instead of simply adding them up may affect the classification performances. Besides, the performance of different clustering algorithms should be evaluated for the CRRBF kernel.